\documentclass[draft]{agujournal2019}
\usepackage{url} 
\usepackage{lineno}
\usepackage[inline]{trackchanges} 
\usepackage{soul}
\draftfalse

\journalname{Geophysical Research Letters}

\begin{document}

%
%


\title{Physics-informed deep learning quantifies propagated uncertainty in seismic structure and hypocenter determination}

%
%




\authors{Ryoichiro Agata\affil{1}, Kazuya Shiraishi\affil{1}, Gou Fujie\affil{1}}


\affiliation{1}{Japan Agency for Marine-Earth Science and Technology, 3173-25, Showa-machi, Kanazawa-ku, Yokohama, Kanagawa 2360001, Japan}




\correspondingauthor{Ryoichiro Agata}{agatar@jamstec.go.jp}



\begin{keypoints}
\item Propose a physics-informed deep learning method to evaluate uncertainty propagation from seismic structure to hypocenter determination
\item Quantify uncertainty in hypocenter determination accurately, evaluating the focal depth relative to the plate boundary quantitatively
\item Demonstrate the potential of physics-informed deep learning for various geophysical inverse problems suffering from uncertainty propagation
\end{keypoints}

%
%

%
%


\begin{abstract}
Subsurface seismic velocity structure is essential for earthquake source studies, including hypocenter determination. 
Conventional hypocenter determination methods ignore the inherent uncertainty in seismic velocity structure models, and the impact of this oversight has not been thoroughly investigated.
Here, we address this issue by employing a physics-informed deep learning (PIDL) approach that quantifies uncertainty in seismic velocity structure modeling and its propagation to hypocenter determination by introducing neural network ensembles trained on active seismic survey data, earthquake observation data, and the physical equation of wavefront movement. 
An analysis of an earthquake in southwest Japan using our method revealed that accounting for such uncertainty propagation significantly reduced the bias and uncertainty underestimation in the hypocenter determination, enabling quantitative evaluation of the focal depth relative to the plate boundary. Our results highlight the potential of PIDL for various geophysical inverse problems, such as investigating earthquake source parameters, which inherently suffer from uncertainty propagation.
\end{abstract}

\section*{Plain Language Summary}

The underground seismic velocity structure of the Earth is essential for determining earthquake locations and analyzing how earthquakes happen.
The models of seismic velocity structure is associated with inherent uncertainty due to limitation of data. 
However, conventional earthquake location methods do not consider the uncertainties in these underground structure models. 
This oversight can lead to inaccurate results.
In this study, we use a new approach that combines deep learning with physical laws to better quantify these uncertainties and see how they affect the determination of earthquake locations. 
We trained neural networks using data from active seismic surveys, earthquake observations, and the physics of how waves move through the Earth.
We applied our method to an earthquake in southwest Japan and confirmed that, by considering these uncertainties, we could significantly improve the credibility of determining the earthquake location. 
This leads to a better understanding of the depth of the earthquake location relative to the boundary of two plates conforming the subduction zone. 
Our results show that advanced deep learning techniques with physical laws could be useful for various geophysical estimation problems, like studying earthquake source locations, which are often associated with uncertainty that propagates through multiple stages of analyses.

\section{Introduction}

In seismogenic zones, realistic subsurface seismic velocity structure models provide fundamental information to understand seismic phenomena. In addition to geophysical interpretation of the structures, they also serve as input data for subsequent analyses such as earthquake source inversions (ESIs), including hypocenter location \cite<e.g.,>[]{Lomax2001GJI,Obana2005JGR,Smith2022}, moment tensor \cite<e.g.,>[]{Ramos-Martinez2001BSSA,Liu2004BSSA,Sugioka2012NatGeo} and finite fault \cite<e.g.,>[]{Graves2001JGR,Masterlark2008GRL,Hori2021} of ordinary and slow earthquakes. 
The estimation of such seismic velocity structures has been primarily based on local earthquake tomography using natural earthquakes \cite<e.g.,>[]{Zhao1992JGR,Zhang2003,Yamamoto2014}. 
In addition, high-resolution seismic velocity structure models,  mainly in two-dimensional (2D) profiles, for shallow portion of subduction zones that host megathrust earthquakes have been obtained through travel-time inversion, such as first-arrival travel time tomography \cite<e.g.,>[]{Zelt1992GJI,Zelt1998JGR,Korenaga2000} and full waveform inversion \cite<e.g.,>[]{Tarantola1984Geophysics,Pratt1996GJI,Virieux2009Geophysics} of active-source seismic data. 
Efforts are now emerging to incorporate this information and create three-dimensional (3D) velocity structure models \cite<e.g.,>[]{Lin2010BSSA,Koketsu2012,Nakanishi2018,Matsubara2019Book,Arnulf2022NatGeo}. 
Seismic velocity models enhanced by active-source exploration data are expected to lay the foundation for accurate monitoring of faulting activities through ESIs.

\begin{figure*}
\begin{center}
\begin{small}
\includegraphics[clip, width=16cm, bb = 0 0 921 454]{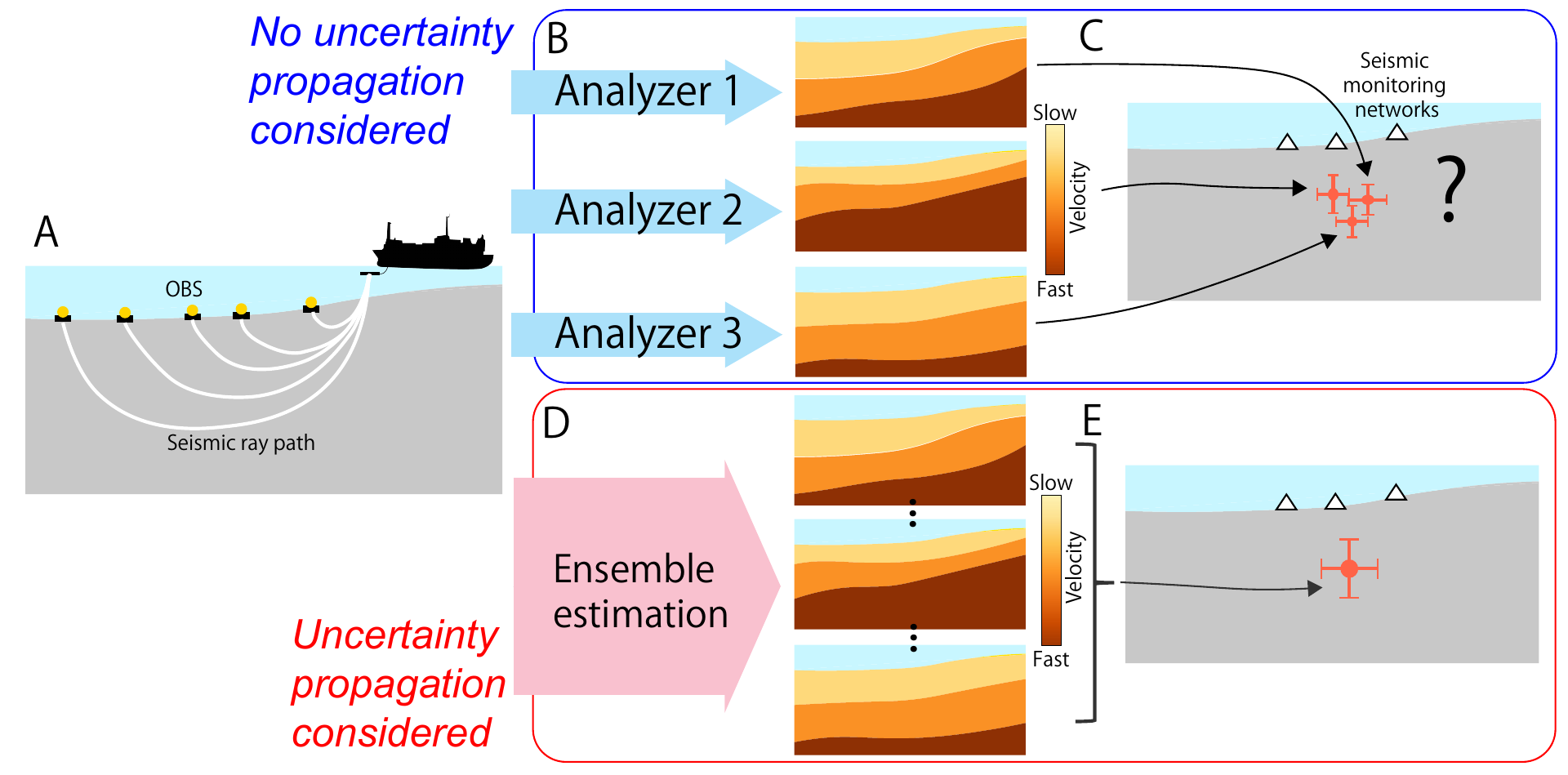}
\end{small}
\end{center}
\caption{
A schematic illustration of the comparison between the proposed (uncertainty propagation considered) and conventional (no uncertainty propagation considered) approaches. 
(A) Example of seismic data acquisition for the estimation of the seismic velocity structure through first-arrival travel time tomography using marine airgun-OBS seismic data. 
The dependence of the seismic ray path on the velocity structure leads to strong nonlinearity in the estimation problem.
Other types of data, such as passive seismic records and ambient noise, can also be used to estimate the velocity structure. 
(B) In conventional methods, the seismic velocity structure models estimated by different individual analyzers using the same or similar seismic datasets may significantly differ owing to the inherent uncertainty in tomographic analyses. 
Moreover, the models may be further simplified by the analyzers.
(C) Schematic illustration of hypocenter determination using the arrival time data from ocean-bottom seismic monitoring networks (white triangles) as an example of ESI using the obtained seismic velocity structure models. 
The orange cross marks schematically correspond to the confidence intervals of hypocenter determination. 
Different seismic velocity structure models may provide different estimations of hypocenter location that are biased by the underestimation of uncertainty. 
In such cases, the extent to which the scientific findings can be trusted is unclear. 
(D) In the proposed method, the uncertainty in velocity structure estimation was quantified by estimating the model as an ensemble. 
(E) By incorporating such a velocity structure ensemble as an input for hypocenter determination, uncertainty propagation is accurately considered and the negative effects of unreasonable model selection and simplification are eliminated.
}
\label{fig:concept}
\end{figure*}

However, careful consideration is required for the dependency between inherent uncertainties in seismic velocity structure estimations and ESIs.
The estimation of the seismic velocity structure is a highly nonlinear inverse problem and is conducted using data limited to the Earth's surface.
These features lead to large uncertainty in the estimation results  at certain depth even in a well-configured 2D estimation using active-source seismic data \cite{Korenaga2000}. 
Due to this uncertainty, the possibility arises for proposing multiple different velocity structure models (Fig. \ref{fig:concept}AB). 
A typical ESI is conducted by selecting a single-velocity structure model such as the average one, often with further model simplification by individual scholars and institutions.
The differences in models assumed by various groups can cause significant variations in estimated source models for the same earthquake, making the reliability of scientific findings unclear \cite{Mai2016} (Fig. \ref{fig:concept}C).
A core of this issue lies in the fact that the propagation of uncertainties from the seismic velocity structure model to ESI is ignored by selecting a single model in advance, resulting in bias and underestimation of uncertainty in ESI \cite{Yagi2011Introduction,Duputel2014,Gesret2015,Agata2021}. 
For accurate and reliable ESI, in addition to quantifying the uncertainties in ESI itself \cite<e.g.,>[]{Tarantola1982,Hirata1987,Lomax2000,Fukuda2008,Minson2014Bayesian}, quantification of uncertainty in the velocity structure estimation and its propagation to ESI must be properly considered. 

Although several efforts have been made to incorporate such uncertainty propagation in ESIs, these were limited to the application of simplified velocity structure models with roughly assumed uncertainty \cite<e.g.,>[]{Duputel2014,Gesret2015,Ragon2018,Agata2022}.  
The impact of propagation of genuine uncertainty inherent in seismic velocity structure estimation to ESIs has not yet been well studied due to technical reasons. 
Existing uncertainty quantification (UQ) methods for seismic velocity structure modeling adopted ensemble-based approaches such as simple Monte Carlo analyses \cite{Zhang1998,Korenaga2000} and Bayesian estimation more recently \cite<e.g.,>[]{Bodin2009,Piana2015,Ryberg2018,Zhang2020Seismic} to cope with inherently strong nonlinearities. 
Difficulty in handling high-dimensional parameter spaces in Bayesian estimation in these existing methods often limits the UQ of seismic velocity structure models to lower-dimensional representations \cite<e.g.,>[]{Bodin2009,Piana2015,Ryberg2018}. 
If realistic samples are obtained, the uncertainty propagation can be evaluated by inputting the ensemble to ESI based on the Bayesian multi-model approach \cite{Gesret2015,Agata2021,Agata2022}. 
However, samples that does not resemble the real velocity structure accurate, as obtained in these methods, make the evaluation of the uncertainty propagation difficult.

The recent remarkable progress in deep learning has led to the emergence of alternative approaches for solving partial differential equations (PDEs) and PDE-based inverse problems, represented by physics-informed neural networks (PINNs) \cite{Raissi2019}. When analyzing observational and experimental data, PINN incorporates information from the physical laws described by the PDE. PINN is particularly advantageous in solving ill-posed inverse problems owing to its inherent regularity \cite{Karniadakis2021}, and has already been applied to several geophysical forward \cite{Smith2021,Waheed2021PINNeik,Song2022GJI,Okazaki2022,Grubas2023,Ding2023EAAI,Ren2024} and inverse problems \cite{Song2021IEEE-TGRS,Izzatullah2022MLST,Rasht-Behesht2022,Fukushima2023JGR} including seismic tomography \cite{Waheed2021PINNtomo,Chen2022JGR}. 
The PINN-based seismic tomography has recently been extended to ensemble-based estimation \cite{Agata2023}. 
In contrast to conventional grid-based approaches, this method yields samples that more closely resemble real velocity structures represented by an infinite-dimensional function, likely leading to accurate quantification of the propagated uncertainty in seismic velocity structure modeling to ESI (Fig. \ref{fig:concept}DE).

Here, we conducted quantification of uncertainty inherent in seismic velocity structure estimation and its propagation to hypocenter determination, the simplest form of ESI, in a real-world problem for the first time using emerging PINN techniques. 
The target event is the Mw 5.9 earthquake that occurred in 2016 in the Nankai Trough region of southwest Japan. 
The experts were initially unsure of whether the earthquake was a plate-boundary earthquake \cite{Wallace2016Mie,Takemura2018}. 
First, we used the first-arrival travel time data obtained from refraction surveys conducted near the hypocenter to conduct UQ for the 2D P-wave velocity structure on the survey line to obtain an ensemble velocity-structure model. 
We then conducted the UQ of hypocenter determination considering the uncertainty propagation from the estimation of velocity structure using this ensemble. 
Moreover, we applied the ensemble to the UQ of the estimation of fault location using seismic reflection survey data collected in the same area. Subsequently, we investigated the fault plane wherein the 2016 earthquake occurred from a statistical viewpoint under properly quantified uncertainty. 
Through these analyses, we found that accounting for such uncertainty propagation from seismic velocity structure estimation significantly reduces the bias and underestimation of uncertainty in the estimated hypocenter location, enabling quantitative evaluation of the focal depth relative to the plate boundary.

\section{Ensemble-based estimation of the P-wave velocity structure}

\begin{figure*}
\begin{center}
\begin{small}
\includegraphics[clip, width=16cm, bb = 0 0 1137 497]{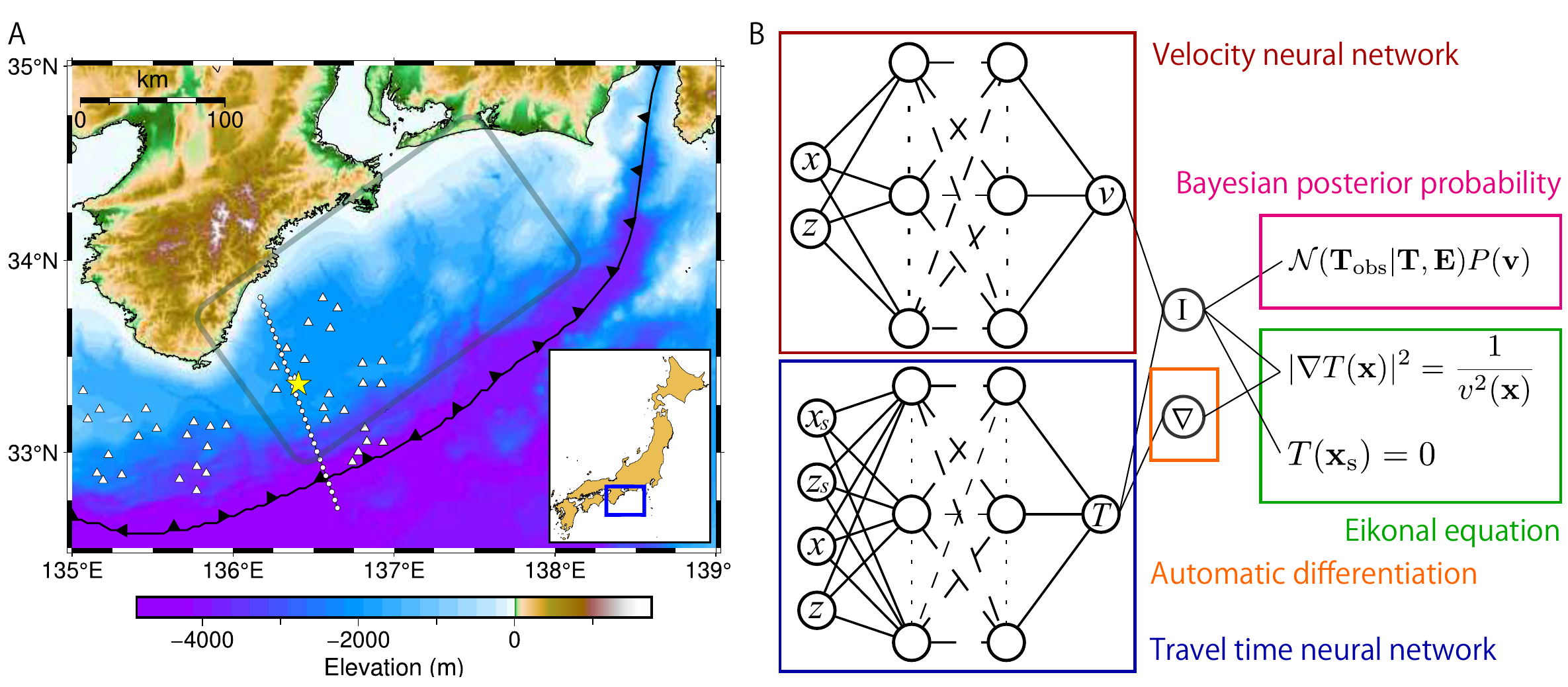}
\end{small}
\end{center}
\caption{
(A) A map of the study area. The yellow star represents the epicenter of the 2016 Mw 5.9 earthquake off the southeastern coast of Mie Prefecture estimated in this study. The circles and triangles represent the locations of OBSs, which were installed in the survey line KI03 and DONET observation nodes, respectively. The gray rectangle represents the approximate focal region of the 1944 Tonankai earthquake. 
(B) A schematic illustration for the NNs of velocity and travel time trained by both the Eikonal equation based on the PINN framework using automatic differentiation and Bayesian posterior probability based on travel time data. 
See Text S1 for the definitions of the mathematical expressions. }
\label{fig:map}
\end{figure*}

The 2016 Mw 5.9 earthquake occurred off the southeastern coast of Mie Prefecture \cite{Wallace2016Mie,Nakano2018Mie,Takemura2018} in the central part of the Nankai Trough region in southwest Japan, which is known for historical megathrust earthquakes \cite{Ando1975}. 
This was the largest earthquake in the region since the 1944 Tonankai earthquake (Mw = 8.0) (Fig. \ref{fig:map}A). 
In the vicinity of the hypocenter, the seismic refraction data acquired using a tuned airgun array and ocean bottom seismometers (OBS) are available along the line KI03.
We estimated the P-wave velocity structure and its uncertainty using 14,146 first-arrival travel times that were manually picked from the refraction data using a PINN-based ensemble estimation method \cite{Agata2023}.
This method represents the seismic velocity structure and travel time function using two different type of neural networks (NNs), respectively, instead of grids or meshes.
The travel time NN is to be optimized for given velocity structure through the PINN framework by minimizing the residual of the Eikonal equation, which can simulate wavefront movement and determine travel time \cite{Smith2021,Waheed2021PINNeik}; the evaluation the residual was conducted in a straightforward manner with the help of automatic differentiation of the NN outputs \cite{Raissi2019}. 
The velocity structure NN gives the velocity structure to this PINN-based training of the travel time NN.  
We generated an ensemble of velocity structure NNs representing the posterior probability for the travel time data formulated by Bayes' theorem, namely, the stochastic property of the estimation uncertainty. 
This was achieved through the combined use of PINN-based travel time calculations and a particle-based variational inference (ParVI) \cite{Liu2016}, which is known to enable more efficient Bayesian estimation than Bayesian sampling method (e.g. Hamiltonian Monte Carlo method \cite{Duane1987}) (Figure \ref{fig:map}B). 
ParVI was performed in the function space inferred by NNs (i.e., function-space ParVI \cite{Wang2019}), not in the space of the weight parameters of NNs. 
As shown in \citeA{Agata2023}, function-space ParVI provides more accurate results than ordinary ParVI when applied to PINN-based Bayesian seismic tomography \cite<e.g.,>[]{Gou2023IEEE-TGRS}.
See Text S1 for improvement of the learning efficiency of high-frequency components \cite{Rahaman2019,Tancik2020,Hennigh2021} and the convergence of ParVI \cite{Scott1979,Xu2022}, the choice of the activation function \cite{Ramachandran2018} and the optimizers \cite{Zaheer2018,Novik2020}, the initialization of NNs \cite{He2015}, and other technical details. 

\begin{figure*}
\begin{center}
\begin{small}
\includegraphics[clip, width=16cm, bb = 0 0 1178 621]{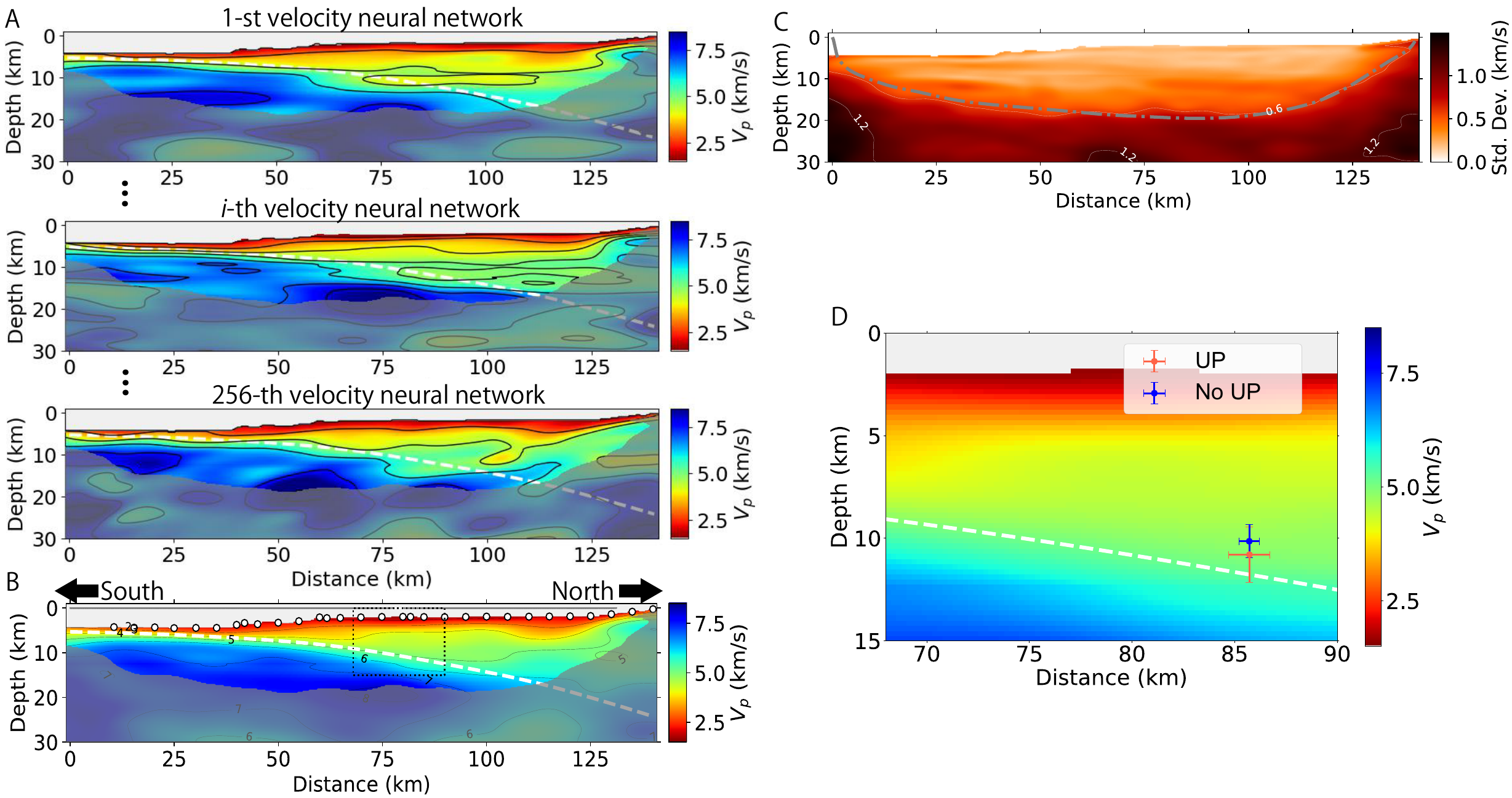}
\end{small}
\end{center}
\caption{
(A) 256 velocity models along the line KI03 represented by NN ensemble members trained by a combination of PINN and ParVI. In (A) and (B), the gray shaded area represents the region with standard deviations larger than 0.6.  
In (A), (B), and (D), the white dashed line represents the plate boundary model proposed by Nakanishi et al.\cite{Nakanishi2018}. 
(B) The mean velocity model calculated based on the estimated ensemble. 
(C) Standard deviation calculated based on the estimated ensemble. The gray dotted-dashed line denotes the bottom of the ray coverage for the mean model.
(D) Comparison between the determined hypocenter locations with and without uncertainty propagation plotted in the region with the dashed rectangle in (B). The cross marks correspond to 2-$\sigma$ intervals. ``UP'' in the legend is the abbreviation of ``uncertainty propagation''. 
}
\label{fig:velocity}
\end{figure*}

From the 256 NNs of velocity obtained through ensemble estimation (Fig. \ref{fig:velocity}A), we obtained the mean model (Fig. \ref{fig:velocity}B) and standard deviation (Fig. \ref{fig:velocity}C) of the seismic velocity structure. 
The obtained mean seismic velocity models clearly show the north-dipping surface of the subducting oceanic plate and low-velocity areas corresponding to an accretionary prism and forearc basin, without the introduction of any prior information. These features are generally in good agreement with existing seismic structures modeled using deterministic tomographic methods \cite{Wallace2016Mie,Nakano2018Mie}.
The standard deviation of the ensemble, namely, the uncertainty of the seismic velocities, was generally low in the area covered by the ray path of the first arrivals. 
This indicated spatial variations even within the ray coverage area, suggesting an uneven distribution of ray paths.

\section{Quantification of propagated uncertainty in hypocenter determination based on the ensemble seismic velocity structure model}

We re-determined the hypocenter of the 2016 earthquake considering the uncertainty propagation from the estimation of the velocity structure using the ensemble estimation results.
We used the P-wave first-arrival time data from nine nearby seismometers installed in the Dense Oceanfloor Network system for Earthquakes and Tsunamis (DONET) \cite{Kaneda2015,Aoi2020} to estimate the source location \cite{Nakano2018Mie}. 
We extended the 2D velocity structure models obtained in the previous process to out-of-plane direction and created 3D volumes, assuming that the spatial variation of the seismic velocity structure in the direction perpendicular to subduction is negligible in the analysis domain \cite{Nakano2018Mie}  (Fig. S1). 
Prior to the determination of the hypocenter, we trained the 256 NNs for 3D travel time function for each ensemble member of seismic velocity structure and input pairs of the source and receiver points sampled from the targeted regions. 
These PINNs surrogate travel time calculation for arbitrary pairs of source and receiver chosen in the region of interest \cite{Smith2022,Taufik2023SciRep} (See Text S2). 
Subsequently, ensemble hypocenter determination was conducted with fast travel time calculations using pre-trained PINNs combined with a ParVI-based Bayesian inversion method \cite{Smith2022}. 
The uncertainty propagation from velocity structure estimation was considered by integrating 256 velocity models into this hypocenter determination method.
This integration was achieved using the Bayesian multi-model approach, a framework that incorporates multiple candidate models in Bayesian inversion \cite{Raftery1997}, which has later been applied to ESIs \cite{Gesret2015,Agata2021,Agata2022}.
See Text S2 for the setting of the likelihood function \cite{Ryberg2019} and the optimizer \cite{Kingma2015}, the choice of hyperparameters \cite{Bishop2006,Watanabe2013,Sato2022}, and other details. 
A total of 128 hypocenter parameter instances (particles) were used to represent the stochastic property of the uncertainty (Fig. S2). 

The mean locations of the hypocenter obtained considering the uncertainty propagation were at 136.41$^\circ$E and 33.36$^\circ$N, and at a depth of 10.81\,km. 
The standard deviations in the horizontal direction along the cross section, vertical direction, and the other horizontal direction were 0.50, 0.68, and 0.38\,km, respectively. These values were larger than those estimated without considering uncertainty propagation, wherein the uncertainty of the seismic velocity structure model is not considered (see Text S3) (Fig. \ref{fig:velocity}D). 
The depth obtained without considering uncertainty propagation is reasonably close to that obtained by a previous model with similar analysis conditions \cite{Nakano2018Mie} (Fig. S3). 

\section{Comparison of the locations of the hypocenter and possible coseismic fault under uncertainty}

In the target region, two structural interfaces exist near the depth of the plate boundary: a megasplay fault and the top of the oceanic crust. At depth, a plate-boundary earthquake is believed to have ruptured the top surface of the oceanic crust. However, at shallow depths, a plate-boundary earthquake possibly does not rupture the top of the oceanic crust, but ruptures the megasplay fault that branches off from it \cite{Park2002,Baba2006,Moore2007}.
Here, we consider both structural interfaces as candidates for the ``coseismic plate boundary \cite{Bangs2009},'' namely, the fault plane ruptured by the plate-boundary earthquake in this region.
We compared our hypocenter with these structural interfaces by considering the uncertainty in their depth related to the seismic velocity structure.
Thus, we re-examined whether this earthquake ruptured a coseismic plate boundary and further investigated whether it was located at the megasplay fault, at the top of the oceanic crust, or elsewhere.
A seismic section in the two-way travel times (TWTs) is available for the survey line TK5, which mostly overlaps with line KI03 \cite{Nakano2018Mie}.
See Text S4 for the details of processing of the seismic data \cite{Yilmaz2001}. 
We manually selected reflectors that likely corresponded to a megasplay fault and the top of the oceanic crust (Fig. S4).
We obtained seismic sections at depth with uncertainty by converting them to TWT using 256 velocity structure models (see Text S4). 
The standard deviations of the depths calculated based on the ensemble were 0.3-0.4\,km for both reflectors. 

\begin{figure*}
\begin{center}
\begin{small}
\includegraphics[clip, width=16cm, bb = 0 0 1639 587]{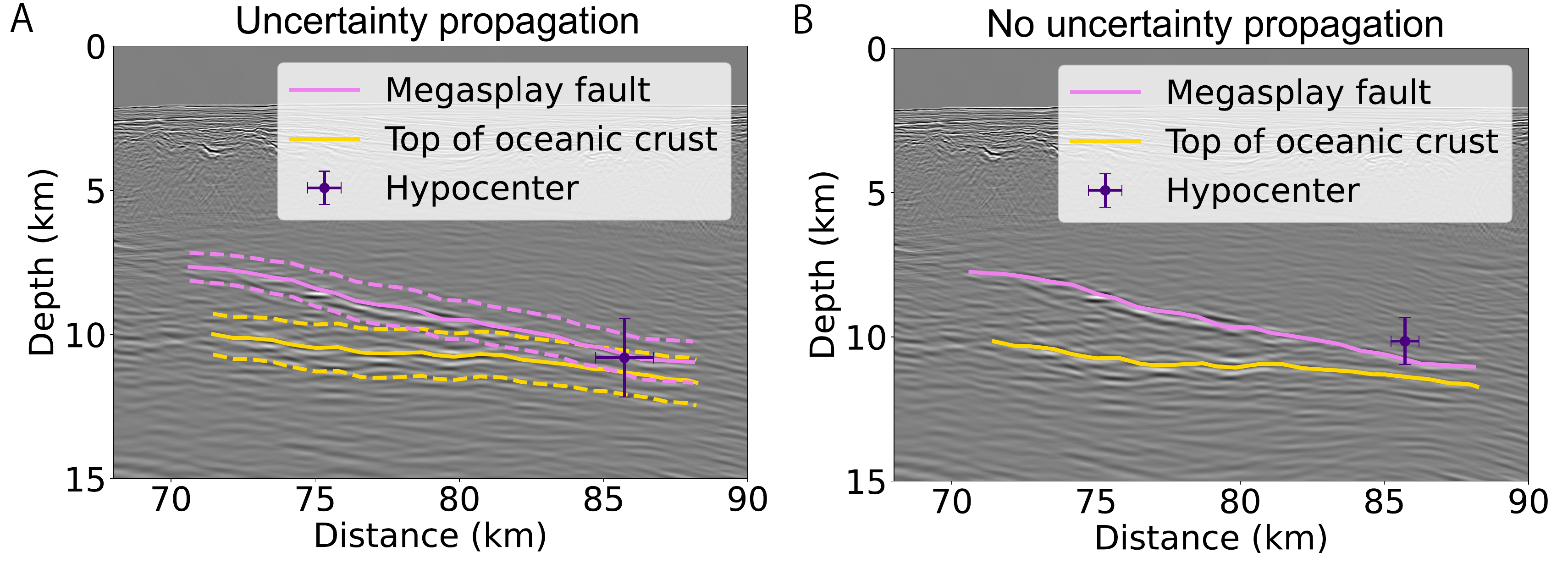}
\end{small}
\end{center}
\caption{A comparison of the estimated hypocenter and fault locations at depth in the region indicated by the dashed rectangle in Fig. \ref{fig:velocity}B. (A) Those obtained using the proposed method considering the uncertainty propagation from the seismic velocity structure model. The dashed lines and crosses correspond to the 2-$\sigma$ intervals of the fault locations and hypocenter, respectively. The background image is the seismic section at depth converted from TWT using the mean velocity model. (B) The results obtained using an ordinary method without considering the the uncertainty propagation. In this case, the two structural interfaces are denoted by single lines.}
\label{fig:reflection}
\end{figure*}

Combining all our obtained results enabled the comparative analysis of the locations of the hypocenter and fault under UQ for the first time. 
The mean location of the megasplay fault agreed well with that of hypocenter depth, and the 2-$\sigma$ intervals largely overlapped between the hypocenter depth and two fault locations (Fig. \ref{fig:reflection}A). 
When an earthquake hypocenter is located at either of the two candidate faults or between them, we assumed that the event occurred at the coseismic plate boundary.
By leveraging the estimated stochastic properties of the uncertainties, we calculated the probability for this case using a one-dimensional normal distribution fitted to the ensemble assuming a finite fault thickness \cite{Rowe2013} (see Fig. S5 and Text S5).
The probability was calculated as 35\%. 
We conducted the same analysis for the case without considering the uncertainty propagation from the seismic velocity structure model. 
In this case, the candidate fault locations were represented by single lines (Fig. \ref{fig:reflection}B). 
As previously indicated, the hypocenter estimate is biased toward being shallower with the underestimation of uncertainty.
The line of the megasplay fault barely overlaps the 2-$\sigma$ interval of the hypocenter, whereas that of the top of the oceanic crust is distant from the interval. 
In this case, the corresponding probability is only 8\%.

\section{Discussion}

In the proposed approach accounting for the uncertainty propagation from the seismic velocity structure model to hypocenter determination, the probability of the occurrence of an event at the coseismic plate boundary was 35\%. 
This is significantly higher than 8\%, which is the probability for the case without considering the uncertainty propagation. 
Previously, the determination of whether the earthquake was at a coseismic plate boundary was based on the latter type of results. 
This requires an expert's evaluation to consider implicit uncertainties that were not quantitatively considered in the analysis; for example, uncertainty of the velocity structure estimation, effect of the simplification of the velocity structure, and uncertainty of the fault location. 
In contrast, the proposed method considering the uncertainty propagation allows us to draw insights solely based on the quantified uncertainty in the analysis results, indicating the potential to achieve a more quantitative and objective characterization of various fault activities in seismogenic zones.

In the results of the proposed approach, the mean depths of the hypocenters and the two fault locations were similar; thus, the shapes of the probability distributions were similar. 
Therefore, we expect that improvements in the accuracy and precision of hypocenter determination, fault location estimation, and velocity structure estimation will further increase the probability. 
For this purpose, the use of the first-arrival time data from other DONET stations distributed over a wider region will provide additional constraints. 
Since 2019, the Japan Meteorological Agency is expected to issue ``Nankai Trough Earthquake Extra Information'' to alert the public when the possibility of the occurrence of a megaquake in the Nankai Trough is assessed to be higher than usual \cite{CabinetOffice2019}. 
When an earthquake of a certain magnitude occurred in the vicinity of the assumed focal region, whether or not the earthquake occurred at the coseismic plate boundary was the basis for additional information. 
DONET provides real-time seismic observations at the ocean bottom of the Nankai Trough region as dense arrays. 
Additionally, rich survey data on the subsurface structures in this region have been compiled in the last two decades \cite{Nakamura2022}. 
Combined with such abundant data, the proposed method considering the uncertainty propagation is expected to provide quantitative and objective information that is highly useful for both the scientific community and public.
This requires extension of the proposed method to estimation of 3D models as the effect of material heterogeneity in the trench-parallel direction cannot be neglected as the target region becomes wider \cite{Shiraishi2019}.

Additionally, we calculated the probabilities of the occurrence of an earthquake at the megasplay fault and the top of the oceanic crust, whose values were 5 and 4\%, respectively. 
These low probabilities were largely attributed to the large uncertainty in the estimated hypocenter depth for the thin fault plane. 
These probabilities also suggest the difficultly in indicating the part of the coseimic plate boundary at which the event occurred based on depth estimates for the dataset we used.
Whether or not a large earthquake ruptures a megasplay fault is an important question as seafloor deformation owing to the rupture of megasplay faults may cause larger tsunamis \cite{Fukao1979}. Although the observation data of the 1944 Tonankai earthquake did not possess sufficient resolution to determine which fault was ruptured, the tsunami inversion result \cite{Baba2006} and some interpretations of nearby subsurface structural information \cite{Park2002,Moore2007} provide supporting evidence of the slip along the megasplay fault. 
However, no conclusions have been reached. 
Further improvements in the precision and accuracy of the involved estimations may provide key information to conclude these arguments. 

Local earthquake tomography, which simultaneously estimates hypocenter locations seismic wave velocity structures, is a well-known method to improve accuracy of hypocenter determination \cite<e.g.,>[]{Zhang2003,Yamamoto2014,Piana2015}.
Unlike local earthquake tomography, the approach taken in this study first determines (or improves) the seismic velocity structure model using tomography analysis of active source seismic data and then performs hypocenter determination based on the model. 
This sequential method is a common approach to incorporate information from active source seismic data into hypocenter determination \cite<e.g.,>[]{Lomax2001GJI,Obana2005JGR}. 
Our study focuses on the uncertainty and its propagation that are often overlooked in this process. 
Both local earthquake tomography and the proposed method are effective in reducing the bias in hypocenter determination results caused by errors in the predetermined seismic velocity structure model. 
Additionally, the proposed method minimizes the dependency on the initial model of the seismic velocity structure, which is a common issue in local earthquake tomography. 
It also provides significanly higher resolution of shallow structures. 
These aspects of the proposed method further contribute to reducing biases and uncertainties in hypocenter determination. 
Recently, efforts have emerged to integrate data from natural earthquakes and active source data in tomography analyses \cite{Arnulf2022NatGeo}. 
Combining the quantification of propagated uncertainties, as considered in this study, with these efforts will become a crucial next step in the future.

We here emphasis two major impacts of introducing NN-based techniques in active-source seismic tomography and hypocenter determination considering uncertainty propagation. 
Firstly, NNs can represent seismic velocity structures in an infinite-dimensional space. 
This representation agrees better with the nature of the real subsurface structure than with the low-dimensional representation often incorporated in conventional ensemble-based velocity structure estimation methods with grid- or mesh-based parameterization \cite<e.g.,>[]{Bodin2009,Piana2015,Ryberg2018}. 
Such an infinite-dimensional representation of the seismic velocity structure enables the accurate and stable subsequent calculations of wavefront movement and conversions of fault depth, leading to an accurate evaluation of uncertainty propagation. 
Introducing NN representation of the physical properties into UQ may have great potential for resolving inverse problems involving significant and complex uncertainty propagation.
Secondly, the determination of the hypocenter based on the Bayesian multi-model framework required only approximately 11 min of computation, even though the travel time for 128 hypocenter ensembles was evaluated for 256 seismic velocity structures in 500 steps. 
Such fast computation was achieved by travel time calculations that leveraged the pre-trained PINNs, which require only forward propagation calculations of NNs. 
The required computation time for the same analysis based on conventional numerical methods such as the fast marching method \cite{Sethian1996} is at least two orders of magnitude higher than that in our approach as a new numerical calculation is required for each instance of the source points and seismic velocity structures. 
Thus, PINN enabled the fast calculation of the hypocenter determination considering the uncertainty propagation; this is particularly beneficial when quick dissemination of information regarding earthquakes to the public is essential, as in the case of Nankai Trough. 
These benefits of deep learning approaches are striking examples of how scientific machine learning \cite{Baker2019} methods, such as PINN, can lead to new scientific breakthroughs.

The concept of the uncertainty propagation-based approach can be extended in a straightforward manner from the determination of hypocenter to ESIs in general, including the rupture process and fault slip inversions for coseismic, aseismic, and interseismic faulting activities using seismic waveforms, crustal deformation, and tsunami data.
For now, the PINN-based approach in geophysical studies has been most successful in eikonal-based hypocenter determination and tomography, thanks to the simplicity of the formulation and the boundary conditions associated with the equation. 
Further development of PINN-based approaches to solve PDEs for elastic waves \cite{Ren2024} and elastic deformation owing to dislocation sources \cite{Okazaki2022} will help the expansion of the proposed strategies to other classes of ESI problems.
Such improvements can enhance ESI and various geophysical inverse analyses, which inherently suffer from uncertainty propagation. 
These advancements are expected to be achieved through the continued  development of PINN and relevant scientific machine learning methods as key actors in the ``AI for Science'' paradigm \cite{Wang2023}. 


\clearpage

\clearpage

\section*{Open Research Section}

The seismic survey data in the line KI03 and TK5 is available from JAMSTEC (Japan Agency for Marine-Earth Science and Technology) Seismic Survey Database \cite{JAMSTEC2004}.

\acknowledgments
This research was supported by JSPS KAKENHI Grant 21K14024 in Grant-in-Aid for Early-Career Scientists. 
Seismic data from JAMSTEC Seismic Survey Database \cite{JAMSTEC2004} were used. 
The seismic survey in the line KI03 performed in 2011 is part of a project entitled ``Research on Evaluation of Linkage between Tokai/Tonankai/Nankai Earthquakes,'' funded by the Japanese Ministry of Education, Culture, Sports, Science, and Technology (MEXT). 
The bathymetric data in the Data and Sample Research System for Whole Cruise Information (DARWIN) in JAMSTEC \cite{JAMSTEC2016} were used. 
Computational resources of the Earth Simulator 4 provided by JAMSTEC was used. 
Some figures were produced using PyGMT \cite{Uieda2021}, namely, a Pythonic interface for Generic Mapping Tools 6 (GMT6) \cite{Wessel2019}
The picked P-wave arrival records of seismic survey data were provided by Dr. Ayako Nakanishi. 
The picked P-wave arrival records of the 2016 Mw 5.9 earthquake off the southeastern coast of Mie Prefecture from DONET seismometers were provided by Dr. Masaru Nakano. 
Comments from Dr. Daisuke Sato and Dr. Hori Takane were valuable for improving the analysis settings and the discussion part in the manuscript, respectively.

%
%


%
%
%
%
%

\end{document}